# Hardware realization of the multiply and accumulate operation on radio-frequency signals with magnetic tunnel junctions

Nathan Leroux[1], Alice Mizrahi[1*], Danijela Marković[1], Dédalo Sanz-Hernández[1], Juan Trastoy[1], Paolo Bortolotti[1], Leandro Martins[2], Alex Jenkins[2], Ricardo Ferreira[2], and Julie Grollier[1]

[1]Unité Mixte de Physique CNRS, Thales, Université Paris-Saclay, 91767 Palaiseau, France

[2]International Iberian Nanotechnology Laboratory (INL), 4715-31 Braga, Portugal

*alice.mizrahi@thalesgroup.com

Artificial neural networks are a valuable tool for radio-frequency (RF) signal classification in many applications, but digitization of analog signals and the use of general purpose hardware non-optimized for training make the process slow and energetically costly. Recent theoretical work has proposed to use nano-devices called magnetic tunnel junctions, which exhibit intrinsic RF dynamics, to implement in hardware the Multiply and Accumulate (MAC) operation – a key building block of neural networks – directly using analogue RF signals. In this article, we experimentally demonstrate that a magnetic tunnel junction can perform multiplication of RF powers, with tunable positive and negative synaptic weights. Using two magnetic tunnel junctions connected in series we demonstrate the MAC operation and use it for classification of RF signals. These results open the path to embedded systems capable of analyzing RF signals with neural networks directly after the antenna, at low power cost and high speed.

**Introduction**

Radio-frequency signals are widely used to convey information and there is an ever increasing demand for their accurate and quick analysis. Recently, artificial neural networks have proven successful for many RF processing applications such as signal classification[1], medical diagnosis[2–4], RF fingerprinting[5], gesture sensing[6], radar applications[7] or aerial vehicle detection and identification[8]. Furthermore, they perform better than conventional algorithms, which rely on feature extractors and complex analysis tools, in terms of resilience to real-world conditions (noisy electromagnetic environment, imperfect RF components or antennas, etc.)[1]. However, processing RF signals with neural networks requires analog signal digitization and training on CMOS-based hardware such as Graphical Processing Units, Tensor Processing Units or dedicated Application Specific Integrated Circuits. This process leads to millisecond delays and consumes hundreds of Watts, limiting the deployment of RF oriented artificial intelligence on embedded systems[9,10].

The need for compact and energy efficient RF hardware has stimulated research on emerging technologies which exhibit native RF dynamics. In particular, spintronic nano-oscillators can receive, transform and emit RF signals[11–13]. They have been proposed as RF emitters and receivers for on-chip communication[14–16] and can implement key operations for RF signal processing, such as spectral analysis[17,18]. Furthermore, they are promising building blocks for neuromorphic computing[19,20]. They have been implemented as neurons in feed-forward neural networks[21], reservoir computing[22–24] and synchronization-based computing[25–27]. Recently, Leroux et al.[28] have proposed using magnetic tunnel junctions as RF synapses to perform the Multiply and Accumulate (MAC) operation, i.e., a weighted sum in hardware, directly on analogue RF signals. The MAC operation is a key operation of neural networks, but it is particularly sensitive to the Von Neumann bottleneck because of the necessity to access memory for reading and writing the weights. Using emerging technologies to perform the MAC operation with collocation of memory and computing is thus projected to offer a hundred fold improvement in energy efficiency[29–34]. Magnetic tunnel junctions can be scaled down to tens of nanometers wide[35] and monolithically integrated into CMOS circuits[36]. Thus, their use as RF synapses provides a path for compact, fast and energy efficient classification of RF signals. Leroux et al. used analytical work and numerical simulations to demonstrate that this MAC implementation could perform classification of handwritten digits images encoded into the RF signals powers[28].

In this work, we experimentally demonstrate that magnetic tunnel junctions can be used as RF synapses with tunable positive and negative weights. We further demonstrate experimentally the MAC operation on two RF inputs with two synapses, and use this MAC to perform 2D classification of RF signals.

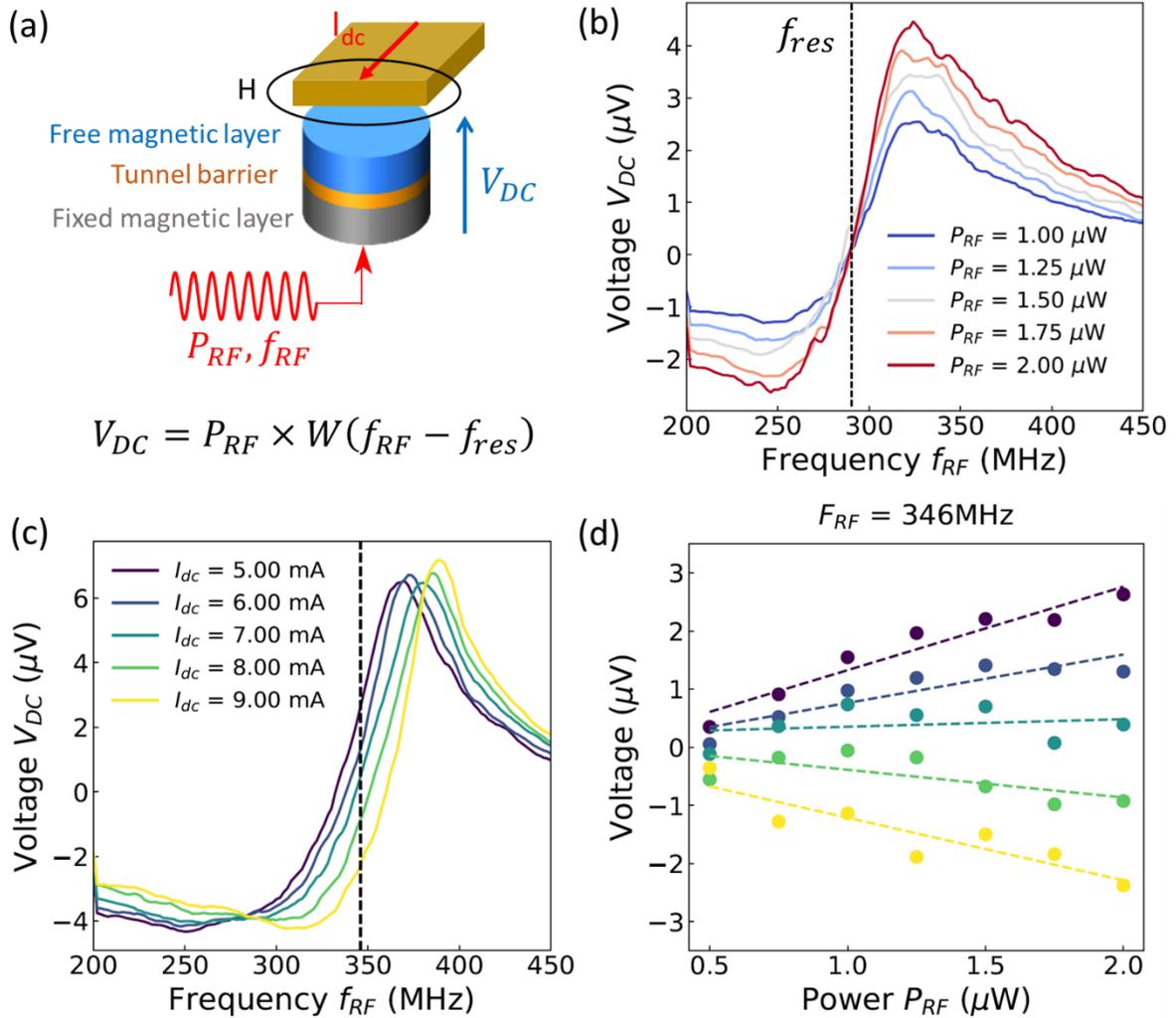

*Figure 1 (a) Schematic of a magnetic tunnel junction performing a multiplication operation on an RF power $P_{RF}$. The synaptic weight is controlled through the resonance frequency $f_{res}$ by application of a local magnetic field via direct current injection in a strip-line placed above the device. (b) Rectified dc voltage generated by the magnetic tunnel junction versus the frequency of the input RF signal, for various input powers. A vertical dashed line indicates the resonance frequency of the magnetic tunnel junction (c) Rectified dc voltage generated by the magnetic tunnel junction versus the frequency of the input RF signal, for an input power of 2 µW and for various dc currents applied in the strip line. The vertical dashed line marks the frequency 346 MHz. (d) Rectified dc voltage versus the input power for an input frequency of 346 MHz and for various dc currents applied in the strip line. All measurements were taken with a 6250 Oe perpendicular magnetic field applied to the device.*

**Methods: using a single magnetic tunnel junction as an RF synapse**

We first demonstrate that a single magnetic tunnel junction can act as a synapse and multiply an RF power by a tunable weight, as schematized in Figure 1 (a). The device is a nanopillar magnetic tunnel junction stack of 250 nm diameter, made of SiO2 // 5 Ta / 50 CuN / 5 Ta / 50 CuN / 5 Ta / 5 Ru / 6 IrMn / 2.0 $Co_{70}Fe_{30}$ / 0.7 Ru / 2.6 $Co_{40}Fe_{40}B_{20}$ / MgO / 2.0 $Co_{40}Fe_{40}B_{20}$ / 0.5 Ta / 7 NiFe / 10 Ta / 30 CuN / 7 Ru, where thickness are indicated in nm. It consists of a reference ferromagnetic layer (2.6 nm Co40Fe40B20), a tunnel barrier (MgO with RA of 8 Ohm µm2) and a free magnetic layer (2.0 nm Co40Fe40B20/ 0.5nmTa / 7nmNiFe, where the CoFeB layer is there to ensure good crystallisation and is fully coupled to the NiFe layer such that they can be considered a single layer). The magnetization of the free layer is in a vortex state. When we inject an RF current through a magnetic tunnel junction, the core of the magnetic vortex in the free layer can be driven into gyrotropic motion, provided the input frequency is near the resonance frequency of the vortex. The tunnel magnetoresistance effect translates these magnetic oscillations into resistance oscillations. The mixing of the input RF current with the resistance oscillations at the same frequency give rise to a dc voltage across the junction, called spin-diode voltage[12,37]. Figure 1 (b) presents this rectified voltage versus the input frequency $f_{RF}$. We observe a caracteristic resonance curve around the resonance frequency (marked by a dashed vertical line). The amplitude of the voltage increases proportionally to the input power $P_{RF}$. By injecting a direct current in the strip-line placed above the device, we can create a local magnetic field collinear to the easy axis of the junction, as schematized in Figure 1 (a). This field modifies the resonance frequency of the vortex, as can be observed in Figure 1 (c), shifting the resonance curve. To demonstrate the synaptic operation, we fix the input frequency $f_{RF}$ at 346 MHz (marked by a dashed vertical line in Figure 1 (c)), which is within the resonance window of the magnetic tunnel junction. Figure 1(d) presents the output voltage $V_{DC}$ versus the input power $P_{RF}$, for various dc currents in the strip-line. We observe that the measured voltage (dots) shows a linear dependence (dashed lines) on the power $P_{RF}$, with the proportionality factor dependent on the dc tuning current, i.e. on the resonance frequency. As a consequence, the magnetic tunnel junction performs the operation:

$$V_{DC} = P_{RF} \times W$$

Where $W = W(f_{RF} - f_{res})$ is the synaptic weight, which is a function of the difference between the resonance frequency and the input frequency. Tuning the resonance frequency allows us to tune the synaptic weight. In order to succesfully implement the synaptic operation, it is critical to be in the low power linear regime where the resonance frequency does not vary significantly with the input power, which is the case here. In future implementations, the tuning of the weight could be achieved in a non-volatile fashion by placing a memristive device on top of the free magnetic layer and electrically controlling its resonance frequency[38]. Note that a single device is sufficient to implement both negative and positive values of the weight, which is an advantage over other synaptic nanodevices[30].

**Results: Demonstration of the MAC operation with a synaptic chain**

In order to demonstrate the MAC operation, we connect by wire bonding two magnetic tunnel junctions in series (MTJ 1 and MTJ 2), forming a synaptic chain, as schematized in Figure 2 (b). Figure 2 shows the characterization of the synaptic chain submitted to a single RF signal. In Figure 2 (a) we plot the voltage across the chain versus the input frequency $f_{RF}$ and observe two resonance peaks, indicated by yellow and green zones, corresponding to MTJ 1 and MTJ 2 respectively. Figure 2 (c) and (d) show the effect of applying dc tuning currents $I_1$ and $I_2$ in the strip lines above MTJ 1 and MTJ 2 respectively. We observe that the resonance frequencies of both MTJs can be tuned within intervals larger than a hundred MHz. Figure 2 (e) and (f) show the voltage across the chain at fixed input frequencies $f^1_{RF}$ = 174 MHz and $f^2_{RF}$ = 540 MHz respectively (dashed lines in Figure 2 (c) and (d) respectively) versus the input power, for different dc tuning currents. Fixing the frequency at $f^1_{RF}$ = 174 MHz (resp. $f^2_{RF}$ = 540 MHz) make it possible to address MTJ 1 (resp. MTJ 2). We observe that both MTJs perform the expected synaptic multiplication: the measured voltage (dots) can be described by a linear model (dashed lines). The synaptic weights $W_1$ and $W_2$ (slopes of the fits) are color encoded and depend on the tuning currents $I_1$ and $I_2$.

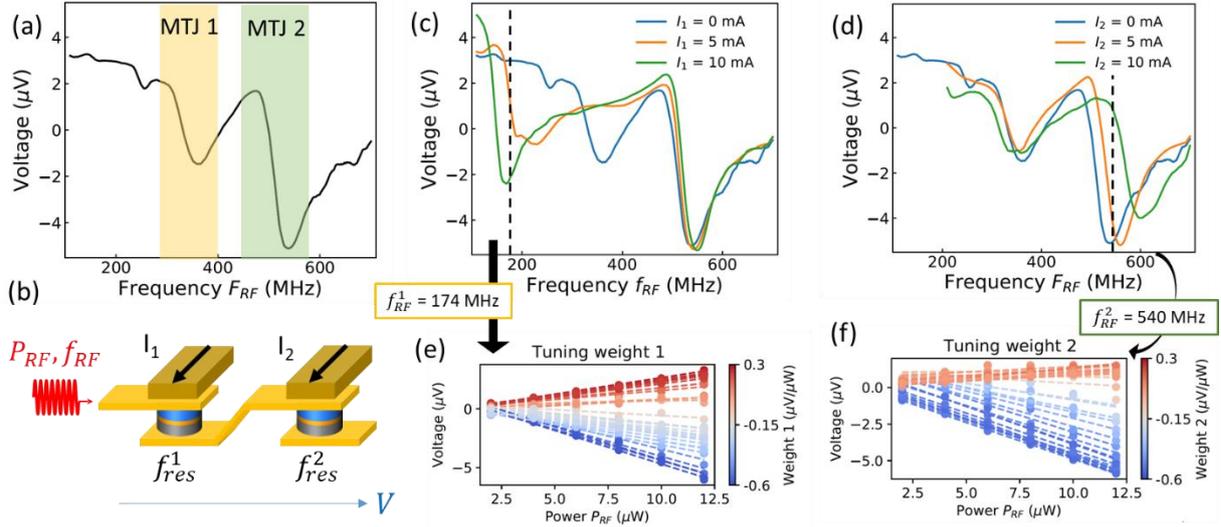

*Figure 2 (a) Rectified dc voltage generated by the chain of two magnetic tunnel junctions versus the input frequency, for an input power of 12 µW. The yellow and green zones qualitatively indicate the frequency windows of MTJ 1 and MTJ 2 respectively. The voltage was smoothed with a moving average of ten points. (b) Schematic of the chain of two magnetic tunnel junctions submitted to an RF signal. (c-d) Rectified dc voltage versus the input frequency, for an input power of 12 µW, and for various values of $I_1$ while $I_2$ = 0 mA (c) and for various values of $I_2$ while $I_1$ = 0 mA (d). The vertical dashed lines mark the frequencies $f^1_{RF}$ =174 MHz (c) and $f^2_{RF}$ = 540 MHz (d). (e-f) Rectified voltage versus input power, for an input frequency of $f^1_{RF}$ =174 MHz (e) and $f^2_{RF}$ = 540 MHz (f), and for various tuning currents in the strip lines above both MTJs. The color bars in (e) and (f) indicate the synaptic weight associated to each MTJ. All measurements were taken with a 3500 Oe perpendicular magnetic field applied to the devices.*

We now test the capability of this chain of two synapses to perform the MAC operation on two RF inputs. The inputs are two RF powers $P_1$ and $P_2$, carried by signals at frequencies $f^1_{RF}$ and $f^2_{RF}$ respectively. We sum the two RF inputs using a power combiner and inject the resulting signal into the synaptic chain, as schematized in Figure 3 (a). We measure the output voltage generated by the chain for different combinations of $P_1$, $P_2$, $W_1$ and $W_2$ (i.e. for different combinations of $P_1$, $P_2$, $I_1$ and $I_2$). $P_1$ and $P_2$ were each varied from 2 to 12 µW by steps of 2 µW while $I_1$ and $I_2$ were each varied from 0 to 10 mA by steps of 2 mA. In Figure 3 (b) we compare the measured voltage across the chain for each combination (blue dots) to the ideally expected MAC voltage computed from single RF signal measurements (black line):

$$V_{MAC} = P_1 \times W_1 + P_2 \times W_2$$

with $W_1 = W_1(f_{RF}^1 - f_{res}^1)$ and $W_2 = W_2(f_{RF}^2 - f_{res}^2)$. For each combination of $I_1$ and $I_2$, the values of $W_1$ and $W_2$ are extracted from the linear fits of the single RF signal input characterization (shown in Figure 2 (e) and (f)). We observe that the experimental voltage matches closely the ideal MAC voltage, with a slope of 0.99 and a root mean square error over all data points of 0.41 µV. In the next section, we evaluate the quality of this MAC operation by using it for RF signals classification then compare it to a simulated noisy MAC.

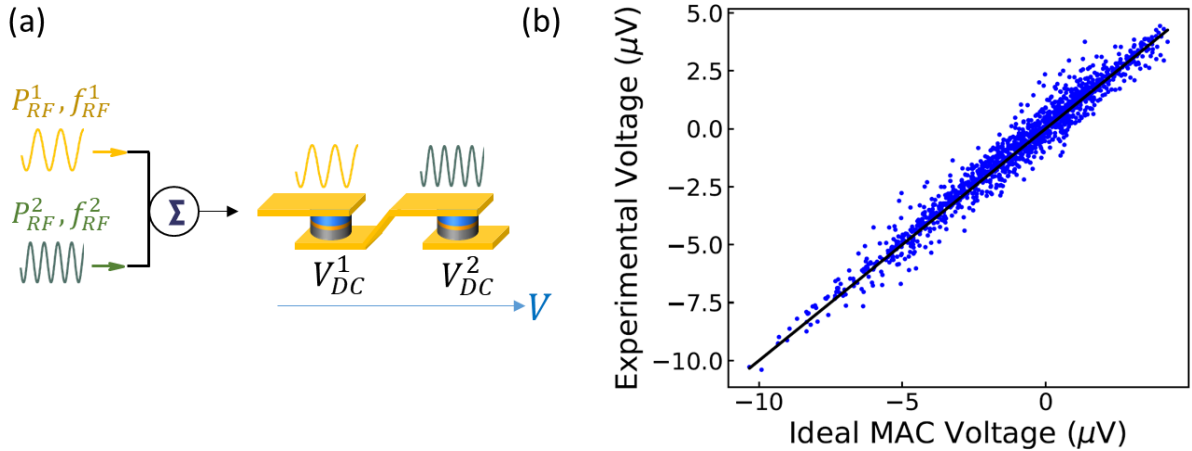

*Figure 3 (a) Schematic of the MAC setup. (b) Measured rectified voltage (blue dots) and ideal MAC voltage (black line) versus the ideal MAC voltage. Each dot corresponds to a ($P_1$, $P_2$, $W_1$, $W_2$) configuration. All measurements were taken with a 3500 Oe perpendicular magnetic field applied to the devices.*

**Discussion: using the hardware RF MAC for RF signal classification**

We illustrate the utility of this MAC operation by a simple classification example, shown in Figure 4 (a). The inputs are the powers $P_1$ and $P_2$ of RF signals at frequencies $f_{RF}^1$ and $f_{RF}^2$. We apply a MAC operation to the inputs using the synaptic chain. If the measured output voltage is positive we classify the ($P_1$, $P_2$) input as "Class 1" (blue squares), while if the voltage is negative we classify the ($P_1$, $P_2$) input as "Class 0" (red dots). Each combination of weights defines a boundary in the 2D plane of axis $P_1$ and $P_2$, and therefore constitutes a possible classification task. Here we first evaluate the classification accuracy for a classification task defined by a fixed predefined combination of weights. Figure 4 (b) shows the classification results for one classification task, corresponding to the following weight configuration: $W_1$ = -0.50 V/W and $W_2$ = 0.23 V/W (achieved by applying $I_1$ = $I_2$ = 0 mA). The dashed line is the boundary between the two classes expected from this weight configuration. The assigned classes are labeled by the blue squares and red dots. For this classification task, the experimental accuracy obtained – i.e. the proportion of correctly classified ($P_1$, $P_2$) inputs – is 100 %.

Figure 4 (c) shows the classification results for all configurations of fixed weights. Each black dot corresponds to a classification tasks defined by a weight configuration. The y-axis is the proportion of correctly classified ($P_1$, $P_2$) inputs while the x-axis is the root mean square error of the measured voltage compared to the ideal MAC voltage, computed over all ($P_1$, $P_2$) inputs. We observe that there is a correlation between classification accuracy and error on the MAC operation: a more precise MAC operation provides more accurate classification. However, we observe a large spread in classification accuracy for each average MAC error. Indeed, the classification accuracy is mainly set by the error brought by inputs close to the boundary (shown in black dashed line in Fig. 4b). This error is itself largely dependent on the values of $P_1$ and $P_2$ (larger powers tend to lead to larger errors). This dependency of errors on inputs explains the observed spread.

The mean accuracy averaged over all classification tasks (weight configurations) of our MAC operation is 93.9 %. In order to better evaluate the suitability of this accuracy, we perform the same classification with a simulated noisy MAC. We use the same inputs and weight configurations as for the experimental MAC. For each ($P_1$, $P_2$, $W_1$, $W_2$) configuration, we simulate the noisy MAC output by a voltage drawn from a normal distribution around the ideal MAC voltage. The standard deviation of the distribution is the mean square error obtained from the experiment for this weight configuration. The red crosses correspond to the classification results for the simulated noisy MAC. The overall accuracy for all the weights configurations is 93.6 %. The fact that the experimental MAC and the simulated noisy MAC provide similar accuracies shows that the experimental MAC does not suffer from systematic errors and is indeed suitable for classification of RF signals.

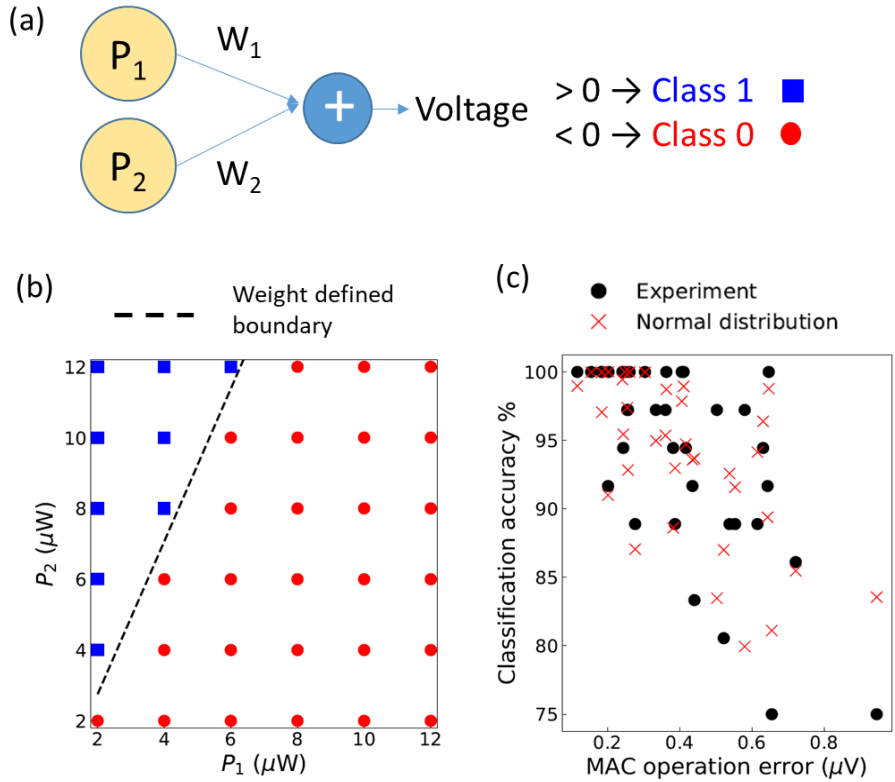

Figure 4 (a) Schematic of the performed 2D classification. A MAC operation with two weights is applied to input ($P_1$, $P_2$). A threshold is applied to the resulting voltage: values above 0 are classified as "Class 1" (blue squares) and values below zero as "Class 0" (red circles). (b) Classification results for all ($P_1$, $P_2$) inputs for one weight configuration: the dashed line represents the expected boundary between the classes 0 and 1, while the blue squares and red circles represent the class of each input as determined by the experimental voltage value. (c) Classification accuracy versus the error to the ideal MAC voltage. Each black circle corresponds to the accuracy of the experimental MAC for a given weight configuration (i.e. a target classification boundary). Each red cross corresponds to the accuracy of the simulated noisy MAC for a given weight configuration (same as the black circles). Each red cross accuracy is computed over all ($P_1$, $P_2$) inputs and average over 100 random trials for each input.

## Conclusion

We have experimentally demonstrated that a magnetic tunnel junction can act as an RF synapse able to perform multiplication of RF powers on analogue RF signals, with tunable synaptic weights taking both positive and negative values. We have also demonstrated that several synapses in series can perform the multiply and accumulate operation, which is at the core of artificial neural networks. We illustrate the application of our system by performing a simple classification task on RF signals. This RF MAC implementation leverages the fact that each synapse is frequency selective, i.e. it only responds to RF signals in its resonance window. We expect this frequency multiplexing scheme to lead to the simplification of architecture for large neural networks, as a single signal combining all inputs can be sent to all synapses of a layer. These results were demonstrated here with vortex-state magnetic tunnel junctions, but could be applied to any spintronic resonator. In order to classify RF signals from various sources for applications, it is critical to have resonators responding to a wide range of frequencies. Vortex-state magnetic tunnel junctions can exhibit frequencies from tens of MHz to several GHz[11], and uniform mode magnetic tunnel junctions or spin Hall oscillators could act as synapses with frequencies up to tens of GHz[26,40]. The same stack of materials can lead to resonators of different frequencies, depending on the size and geometry of each device. Therefore, one could build an array of synapses of different resonance frequencies, able to respond to, and classify, RF signals over the 50 MHz – 50 GHz frequency range in parallel. Magneto-electrical effects can also be used to tune synaptic weights in a non-volatile fashion. This opens the path for compact, fast and energy efficient devices, allowing RF-signal classification in embedded systems.

**Acknowledgements**

This work is supported by the European Research Council ERC under Grant No. bioSPINspired 682955, the French ANR project SPIN-IA (Grant No. ANR-18-ASTR-0015), and the French Ministry of Defense (DGA).